%% file: main_arxiv.tex
\documentclass[longbibliography,showpacs,floatfix,superscriptaddress,twocolumn,aps,prb]{revtex4-2}

\usepackage{graphicx}
\usepackage{amsfonts}
\usepackage{float}
\usepackage{amssymb}
\usepackage{bm}
\usepackage{amsmath}
\usepackage{mathtools}
\usepackage{psfrag}
\usepackage{multirow}
\usepackage{tabularx}
\usepackage{textcomp}
\usepackage{units}
\usepackage{lipsum}
\usepackage{physics}
\usepackage{soul}
\usepackage{titlesec}
\usepackage{times}
\usepackage[colorlinks=true,citecolor=blue,linkcolor=magenta,hypertexnames=false]{hyperref}
\usepackage{enumitem}
\usepackage[dvipsnames]{xcolor}
\usepackage{tikz}

\newcommand{\vhat}[1]{\boldsymbol{\hat{#1}}}
\newcommand{\avg}[1]{\langle #1 \rangle}

\newcommand{\beginsupplement}{%
    \setcounter{equation}{0}
    \renewcommand{\theequation}{S\arabic{equation}}
    \setcounter{section}{0}
    \renewcommand{\thesection}{S\arabic{section}}  
    \setcounter{table}{0}
    \renewcommand{\thetable}{S\arabic{table}}  
    \setcounter{figure}{0}
    \renewcommand{\thefigure}{S\arabic{figure}}
    \renewcommand{\citenumfont}[1]{S##1}
    \renewcommand{\bibnumfmt}[1]{[S##1]}
    \setcounter{page}{1}
}

\begin{document}

\title{Anharmonic Collective Oscillations in Isotropic Spin Systems and their Spectroscopic Signatures}

\author{Anna Fancelli}
\affiliation{Helmholtz-Zentrum Berlin f\"ur Materialien und Energie, Hahn-Meitner-Platz 1, 14109 Berlin, Germany}
\affiliation{Dahlem Center for Complex Quantum Systems and Fachbereich Physik, Freie Universit\"at Berlin, Arnimallee 14, 14195 Berlin, Germany\looseness=-1}
\author{Mat\'ias G. Gonzalez}
\affiliation{Helmholtz-Zentrum Berlin f\"ur Materialien und Energie, Hahn-Meitner-Platz 1, 14109 Berlin, Germany}
\affiliation{Dahlem Center for Complex Quantum Systems and Fachbereich Physik, Freie Universit\"at Berlin, Arnimallee 14, 14195 Berlin, Germany\looseness=-1}
\affiliation{Institute of Physics, University of Bonn, Nussallee 12, 53115 Bonn, Germany}
\author{Subhankar Khatua}
\affiliation{Department of Physics and Astronomy, University of Waterloo, Waterloo, Ontario N2L 3G1, Canada}
\affiliation{Department of Physics, University of Windsor, 401 Sunset Avenue, Windsor, Ontario N9B 3P4, Canada}
\affiliation{
Institute for Theoretical Solid State Physics, IFW Dresden and W\"urzburg-Dresden Cluster of Excellence ct.qmat, Helmholtzstr. 20, 01069 Dresden, Germany}
\author{Bella Lake}
\affiliation{Helmholtz-Zentrum Berlin f\"ur Materialien und Energie, Hahn-Meitner-Platz 1, 14109 Berlin, Germany}
\affiliation{Institut f\"ur Festk\"orperforschung, Technische Universit\"at Berlin, 10623 Berlin, Germany}
\author{ Michel J. P. Gingras}
\affiliation{Department of Physics and Astronomy, University of Waterloo, Waterloo, Ontario N2L 3G1, Canada}
\affiliation{Waterloo Institute for Nanotechnology, University of Waterloo, Waterloo, Ontario N2L 3G1, Canada}
\author{Jeffrey G. Rau}
\affiliation{Department of Physics, University of Windsor, 401 Sunset Avenue, Windsor, Ontario N9B 3P4, Canada}
\author{Johannes Reuther}
\affiliation{Helmholtz-Zentrum Berlin f\"ur Materialien und Energie, Hahn-Meitner-Platz 1, 14109 Berlin, Germany}
\affiliation{Dahlem Center for Complex Quantum Systems and Fachbereich Physik, Freie Universit\"at Berlin, Arnimallee 14, 14195 Berlin, Germany\looseness=-1}

\date{\today}

\begin{abstract}
Spin waves are the fundamental excitations in magnetically ordered spin systems and are ubiquitously observed in magnetic materials. 
However, the standard understanding of spin waves as collective spin oscillations in an effective harmonic potential does not consider the possibility of soft modes, such as those due to an effective {\it quartic} potential. 
In this work, we show that such quartic potentials arise under very general conditions in a broad class of isotropic spin systems without a fine-tuning of the interaction parameters. 
Considering models with spin spiral ground states in two and three spatial dimensions, we numerically demonstrate that quartic amplitude spin oscillations produce a fluctuation-induced spin-wave gap which \emph{grows} with temperature according to a characteristic power-law.
In conjunction with a phenomenological theory, the present work provides a general theoretical framework for describing soft spin modes, extending the previously discussed spin dynamics in the presence of order-by-disorder, and highlighting the important role of finite-size effects. Our predictions of a temperature-dependent gap in spiral spin systems could be tested in inelastic neutron scattering experiments, providing direct spectroscopic evidence for thermal effects arising from soft spin modes in magnetic materials.
\end{abstract}

\maketitle

%\textit{Introduction.} 
\textit{Introduction.}--- The notion of a quasiparticle is widely used to describe collective excitations in many-body systems~\cite{Wolfle2018}. 
This term refers to a common situation in which, similarly to fundamental particles in high-energy physics, a collective excitation in a solid-state system can be described by characteristic properties such as its effective mass or, more generally, its dispersion relation. 
Well-known examples of quasiparticles include phonons, which arise from collective lattice vibrations, and magnons, which result from collective spin precessions~\cite{kittel1991}. 
In both cases, the quasiparticles can be classically understood as oscillations in an effective harmonic potential $V(x)=\alpha x^2$ with a frequency $\omega$, and with the quantum mechanical excitation energy, $E$, given by $E=\hbar\omega$.

In this Letter, we address the fundamental question of how the concept of a quasiparticle changes when the effective potential is softer than quadratic and how such anharmonicity can be exposed in experiments. 
For concreteness, we specifically consider the case of {\it quartic} potentials where $V(x)\propto x^4$. 
Such potentials and the associated collective excitations are of relevance in quite distinct physical systems, including trapped Bose-Einstein condensates~\cite{Fetter2001,Gygi2006}, chaotic systems~\cite{LAKSHMANAN1993,Bannur1997}, soft phonons~\cite{Lan2015,Wehinger2016}, and, in high-energy physics, in the context of the Higgs self-coupling~\cite{Csaki2020,Nima2002,Nima2002_2}. 
Here, we investigate quartic potentials in classical Heisenberg spin systems, which emerge under fairly general conditions and without the need to fine-tune model parameters to eliminate the harmonic  $\alpha x^2$ component.
Our starting point is a Heisenberg spin Hamiltonian that has several degenerate symmetry-related planar spin spiral ground states. 
We show that in a system exhibiting a planar spin-spiral long-range ordered state at one of several symmetry-related ground-state ordering wave vectors, an oscillation of the spin component perpendicular to the spiral ordering plane, taken at the wave vector of a different ground-state spiral, has a quartic dependence of the excitation energy on the amplitude. We refer to this oscillation as a {\it quartic} oscillation in the following.

Since a system with a purely quartic potential has a vanishing harmonic component ($\alpha=0$), one might naively expect it to display gapless excitations.
While this is true in the zero-temperature limit, our numerical simulations reveal that thermal fluctuations dynamically generate a spin-wave gap $\Delta$. This implies a remarkable situation in which the elementary properties of a quasiparticle, such as its effective mass, are no longer determined by the system's microscopic parameters, but instead become state-dependent, allowing them to be modified by external parameters such as the temperature $T$.
Our numerical simulations of finite size spin systems in two and three spatial dimensions indicate a characteristic scaling of this fluctuation-induced spin wave gap of the form $\Delta(T)\sim T^\mu$ where $\mu=1/4$ ($\mu\approx 1/2$) at low (intermediate) temperatures. 
We explain these results using a phenomenological theory that takes into account the quartic oscillations and their coupling to other quadratic modes that inevitably exist in the system. 
Specifically, this theory reveals an intriguing finite-size effect resulting from the small number [$O(1)$] of quartic modes that couple to $O(N)$ quadratic modes (where $N$ is the total number of spins). 
Consequently, although the low-temperature gap scaling with $\mu=1/4$ is a direct consequence of the quartic oscillation, the low-temperature regime in which it occurs vanishes in the thermodynamic limit, leaving only the thermally generated gap with $\mu=1/2$.

Through our detailed understanding of the gap scaling associated with quartic spin oscillations, we extend previous theoretical work on a related type of fluctuation-induced spin-wave gap that has been predicted in the context of a thermal or quantum order-by-disorder effect due to an exact but accidental ground-state zero mode, giving rise to a so-called {\it pseudo-Goldstone mode}~\cite{Rau2018pseudogoldstone, Gohlke2020emergence, Khatua2023PseudoG, Hickey25}.
In the thermal classical case, the spin-wave gap was found to scale with the same characteristic exponent $\mu=1/2$~\cite{Khatua2023PseudoG}. 
This implies that pseudo-Goldstone modes do {\it not} require the existence of exact ground-state zero modes, which are rarely found in real materials. Instead, they can be observed under less stringent conditions in the presence of quartic potentials, which we demonstrate can occur under rather general assumptions for isotropic spin systems.

On the experimental front, the dynamical gap generation atop an anharmonic potential we study may have already been detected in neutron scattering experiments on magnetic materials such as the pyrochlore helimagnet ZnCr$_2$Se$_4$~\cite{Tymoshenko2017PseudoGoldMat, Inosov2020} and the half-Heusler compound GdPtBi~\cite{Sukhanov2020}. 
Here, we provide the theoretical basis for these observations by relating them to quartic oscillations using a classical description to predict the ensuing temperature dependence of the spin-wave gap.

\begin{figure}[t!]
    \centering
    \includegraphics[width=0.9\linewidth]{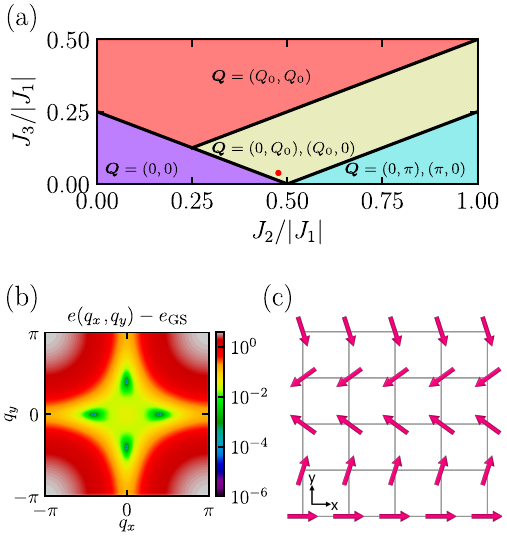}
    \caption{(a) Zero temperature phase diagram for the $J_1$-$J_2$-$J_3$ Heisenberg model on the square lattice, with ferromagnetic $J_1<0$. 
    Four different phases are present: ferromagnetic order with $\bm{Q}=(0,0)$ (purple region), stripe order with $\bm{Q}=(0,\pi),(\pi,0)$ (light blue region), one dimensional spiral with $\bm{Q}=(0,Q_0),(Q_0,0)$ (yellow region), two dimensional spiral with $\bm{Q}=(Q_0,Q_0)$ (red region).
    The red point corresponds to the specific values of $J_2$ and $J_3$ used in this work; panels (b) and (c) refer to this point. (b) Energy difference per spin as a function of the wave vector $\bm{q}$ of the coplanar spiral, $e(q_x,q_y)-e_{\text{GS}}$, where $e_{\text{GS}}$ denotes the energy minimum. The energy minima are at $ \bm{Q} = (0,\pm \frac{2\pi}{5})$ and $\bm{Q}=(\pm \frac{2\pi}{5},0)$. 
    (c) Illustration of the ground state configuration with the spiral along the $\hat{\bm{y}}$ axis. This configuration corresponds to $\bm{Q}_{\rm B}=  (0, \pm \frac{2\pi}{5})$ (the selected order with associated Bragg peak).
    }
    \label{fig:model}
\end{figure}

\textit{Spin Systems and Quartic Perturbations.}--- To set the stage for our study of anharmonic spin oscillations in isotropic spin systems, we first consider a minimal two-dimensional (2D) square lattice model that supports soft modes in a quartic potential. 
Later, we will extend this analysis to a three-dimensional (3D) cubic lattice to confirm that our results apply to a wider range of spin systems and the more common 3D situation. 
In the square lattice case, we consider a Heisenberg Hamiltonian with exchange couplings extending up to third-nearest neighbors, with ferromagnetic $J_1<0$ and antiferromagnetic $J_2$, $J_3>0$ couplings:
\begin{equation}
H=J_1\sum_{{\langle i<j \rangle}_1} \bm{S}_{i}\cdot \bm{S}_{j} + J_2\sum_{{\langle i<j \rangle}_2} \bm{S}_{i}\cdot \bm{S}_{j} + J_3\sum_{{\langle i<j \rangle}_3} \bm{S}_{i}\cdot \bm{S}_{j}.
\label{eq:Ham}
\end{equation}
Here, $\langle\ldots \rangle_n$ indicates the sum over the $n^{\textrm{th}}$-nearest neighbors, and the spins are treated classically as three-dimensional vectors with unitary norm $\lvert {\bm S}_{i} \rvert = 1$. The ground-state spin configuration for Eq.~\eqref{eq:Ham} is a coplanar spiral~\cite{Rastelli19791},
\begin{equation}
\bm{S}^{\text{GS}}_{i} = \begin{pmatrix} \cos(\bm{Q}\cdot \bm{r}_{i}+\phi)  \\ \sin(\bm{Q}\cdot \bm{r}_{i}+\phi) \\ 0 \end{pmatrix},
\label{eq:Qord}
\end{equation}
where $\bm{Q}=(Q_x,Q_y)$ is the spiral wave vector, $\bm{r}_i$ is the position of site $i$, and $\phi\in[0,2\pi)$ is a phase.
Without loss of generality, we assume the spiral to be in the $x$-$y$ plane and also set $\phi=0$.
The wave vector $\bm{Q}$ is then determined by substituting the spiral spin configuration into the Hamiltonian and identifying the $\bm{Q}$ that minimizes the system’s energy. We fix $J_1=-1$, $J_2=0.476393205$, and $J_3=\frac{1}{2}J_2 - \frac{1}{5}$, such that the ground state is a one-dimensional spiral with two possible commensurate wave vectors $ \bm{Q} = (0,\pm \frac{2\pi}{5})$ and $(\pm \frac{2\pi}{5},0)$, related by a $\pi/2$ lattice rotation (see Fig.~\ref{fig:model}). 
While the commensurability of the spiral is irrelevant to the physical properties that we study herein, it tremendously simplifies the numerical calculations as it enables periodic boundary conditions.

The continuous O(3) symmetry of the Hamiltonian cannot be broken at finite temperatures in two dimensions, excluding a transition into a long-range magnetically ordered phase \footnote{The absence of a phase transition in 2D follows from the Mermin-Wagner-Hohenberg theorem~\cite{MerminWagner,Hohenberg}.
Below a certain temperature, finite systems appear magnetically ordered when the correlation length becomes comparable to the system size.
However, true long-range order is absent in the thermodynamic limit.}. 
However, the system undergoes a finite-temperature Ising transition associated with the discrete $\mathbb{Z}_2$ symmetry between the two (energetically degenerate) spiral wave vector directions~\cite{Seabra2016NovelphasesSquare}.
Specifically, the one-dimensional spiral phase breaks the square lattice's fourfold rotational symmetry down to a twofold symmetry, with the spiral oriented along either of the Cartesian directions (see Fig.~\ref{fig:model}).
The selected ordering wave vector, $\bm{Q}_{\rm B}$, is responsible for the Bragg peaks of the ground-state order, and we refer to the wave vector that is not selected as $\bm{Q}_{\rm aB}$, where ``aB'' stands for anti-Bragg.

\begin{figure}[t!]
     %\centering   
     %\hspace*{-1.3 cm}    
     \includegraphics[width=0.87\linewidth]{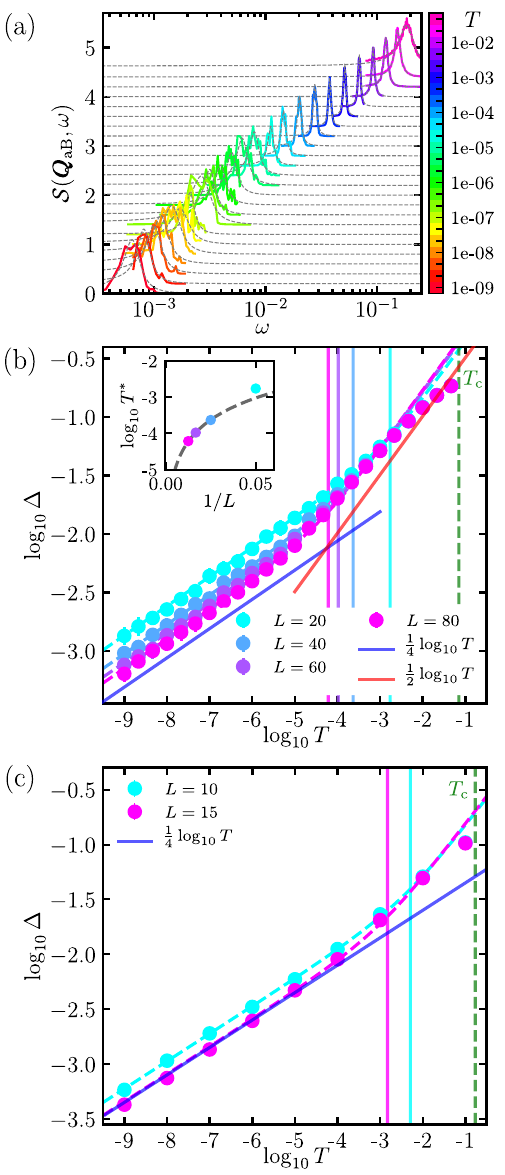}
     \caption{(a) Normalized dynamical structure factor $\mathcal{S}(\bm{q},\omega)$ at $\bm{q}=\bm{Q}_{\rm aB}$ for different temperatures, obtained using sMD simulations for $L=80$ in the 2D case. The gray-dashed lines show Lorentzian function fits. The position of the maximum indicates the gap $\Delta$. (b) Logarithm of the gap $\Delta$ as a function of the logarithm of the temperature in the 2D case. To guide the eye, we plot two reference lines with slopes of 1/4 (blue line) and 1/2 (red line). The colored points and the dashed lines correspond to the data (with error bars defined in the SM~\cite{sup}) and fitting function [Eq. \eqref{eq:omegafun}], respectively. The vertical lines correspond to $ \log_{10}{T^*}$ with $T^*=6\lambda / \alpha^2$, where the fitted function changes slope for the different sizes. The inset shows the finite-size scaling analysis for this quantity. (c) The same information for the 3D model.}
     \label{fig:square_gapT}
\end{figure}

This system has two key properties that are essential in harboring the physics of anharmonic collective excitations that we aim to explore.
First, the system supports spin oscillations in an effective quartic potential which correspond to spin fluctuations in the direction {\it perpendicular}
to the planar spiral ground state in Eq.~(\ref{eq:Qord}) and characterized by the wave vector $\bm{Q}_{\rm aB}$. 
Specifically, we consider $\bm{S}^{\text{quartic}}_{i}$ as a perturbed state about the long-range ordered ground state,
\begin{equation}
\bm{S}^{\text{quartic}}_{i}=\frac{1}{N_i} \left[ \bm{S}^{\text{GS}}_{i} + \delta \begin{pmatrix}  0 \\ 0 \\  \cos(\bm{Q}_{\rm aB}\cdot \bm{r}_{i}+\phi') \end{pmatrix}\right].
\label{eq:quartic}
\end{equation}
Here $\delta$ is the perturbation amplitude, $N_i$ is the site ($i$)-dependent normalization to ensure $\lvert {\bm S}^{\text{quartic}}_{i} \rvert = 1$, and $\phi'\in[0,2\pi)$.  
We show in the Supplemental Material (SM)~\cite{sup} that such a perturbed non-coplanar state always has a vanishing $\delta^2$ term in the energy, regardless of the range of the interactions, as long as the Hamiltonian conserves the $C_4$, reflection, and translation symmetries of the square lattice. 
For our particular set of interactions, the energy per site is $e = e_{\rm GS} + 0.039 \delta^4 + O(\delta^6)$~\cite{sup}. 
Second, aside from three zero modes arising from perturbations with the wave vector $\bm{Q}_{\rm B}$ --- which correspond to global spin rotations that preserve the spiral order and would represent real Goldstone modes at $\bm{Q}_{\rm B}$ were the system magnetically long-range ordered at $T>0$ --- the system does not exhibit any additional modes softer than the quartic ones discussed above. 
In particular, the system has no accidental ground-state degeneracies~\footnote{Accidental continuous ground-state degeneracies are known to occur in systems with fine-tuned interactions that display simpler ground-state wave vectors such as $\bm{Q}=(0,\pm\pi)$ and $\bm{Q}=(\pm\pi,0)$~\cite{Henley1989}. 
By considering systems with spin spiral ground states, we avoid such accidental zero modes.} leading to zero modes, such as occurs in the context of order-by-disorder phenomena~\cite{Khatua2023PseudoG}. 
Therefore, the system considered allows us to focus specifically on the effects of quartic modes, which have heretofore attracted limited attention.

To further establish the general scope and quantitative validity of our findings, we extend our analysis to a 3D  classical Heisenberg model on a cubic lattice. 
As in the 2D case, we consider a model with general spiral ground states. 
This is achieved by using a Hamiltonian with the following interaction parameters: $J_1= -1 $, $J_2=4/[4(4 + 1.8\cos{\gamma})]$, and $J_4=1.8/[4(4 + 1.8\cos{\gamma})]$, where $J_n$ is the coupling between $n$th-nearest neighbors~\cite{Tymoshenko2017PseudoGoldMat}.
In this case, the ground state is also given by a coplanar one-dimensional spiral given by Eq.~\eqref{eq:Qord}, but with ordering wave vectors $\bm{Q}=(0,0,\pm \gamma)$, $(0,\pm \gamma,0)$ or $(\pm \gamma,0,0)$.
We fix $\gamma = 2\pi/5$, yielding again a five-site periodic spiral. 
Below a critical temperature $T_c$, one of these three spiral configurations is selected, and the perpendicular spin fluctuations with the not-selected wave vectors correspond to quartic oscillations~\cite{sup}. 

There are two main differences between this 3D model and the 2D model discussed above. First, quartic oscillations can now involve contributions from \emph{two} wave vectors, corresponding to the not-selected spirals in the ordered state. 
Second, in 3D, the SO(3) spin symmetry is explicitly broken in the ordered phase below $T_c$, giving rise to true magnetic long-range order (we find that both types of symmetry breaking, lattice rotation and spin rotation, occur at the same critical temperature $T_c$).

\textit{Numerical simulations.}--- To investigate the dynamic signatures of quartic modes, we perform classical Monte Carlo (cMC)~\cite{Alzate19} and spin molecular dynamics (sMD)~\cite{Alder57} simulations on the models introduced above. 
Using equilibrated states generated from cMC as initial configurations for the sMD simulations, we compute the spin structure factor, $\mathcal{S}(\bm{Q},\omega)$, as a function of temperature~\cite{sup}. 
Specifically, our focus is on the determination of the excitation gap $\Delta$ at the wave vector $\bm{q}=\bm{Q}_{\rm aB}$ of the quartic perturbation in Eq.~(\ref{eq:quartic}), which can be extracted by fitting the spectra with a Lorentzian function~\cite{sup}. 

In the 2D case, the cMC results show that at $T_c = 0.071(3)$ the system breaks the fourfold lattice rotation symmetry and develops a one-dimensional spiral along the $\hat{\bm{x}}$ or $\hat{\bm{y}}$ direction. 
Below $T_c$, $\mathcal{S}(\bm{Q}_{\rm aB},\omega)$ displays a peak at a finite frequency that decreases with temperature [see Fig.~\ref{fig:square_gapT}(a)].
Extracting the value of the gap for each temperature results in Fig.~\ref{fig:square_gapT}(b), where we show the gap $\Delta$ as a function of the temperature in a log-log plot for several square systems with linear sizes $L=20$, 40, 60, and 80. 
This gap can be attributed to an anharmonic effect because linear spin wave theory would predict a gapless mode at this point as it treats quartic oscillations as zero modes~\cite{Tymoshenko2017PseudoGoldMat}.

In the 3D case, we use cubic systems with $N=L^3$ spins to compute $\mathcal{S}(\bm{Q}_{\rm aB},\omega)$. 
The cMC calculations show that at $T_c \simeq 0.17(1)$, the system orders in a one-dimensional spiral along $\hat{\bm{x}}$, $\hat{\bm{y}}$ or $\hat{\bm{z}}$. 
As expected, the dynamical structure factor at $\bm{q}=\bm{Q}_{\rm aB_1}$ and $\bm{q}=\bm{Q}_{\rm aB_2}$ displays a peak at the same frequency~\cite{sup}. 
The results are shown in Fig.~\ref{fig:square_gapT}(c) for $L=10$ and 15. The similarities between the results for the 2D and 3D models show that the gap behavior is a characteristic property of a system that possesses quartic perturbations.

Overall, our results can be summarized by three key observations, which we will explain in detail below
\begin{enumerate}
    \item At low temperatures, the gap is well described by the scaling $\Delta \sim T^{1/4}$ for both the 2D and 3D models, as can be seen in Fig.~\ref{fig:square_gapT}(b) and \ref{fig:square_gapT}(c), respectively.
    \item At higher temperatures, the gap increases faster than $\Delta \sim T^{1/4}$. 
    Remarkably, and maybe counter-intuitively, the largest gap is found right below the phase transition, although the order nearly vanishes in this regime. This indicates that the gap at $\bm{Q}_{\rm aB}$ is a robust feature of the ordered phase.
    \item Finite-size effects are more prominent at low $T$, where $\Delta \sim T^{1/4}$, as seen in Fig.~\ref{fig:square_gapT}(b) and \ref{fig:square_gapT}(c). 
\end{enumerate}

\textit{The $T^{1/4}$ behavior.}--- To explain the origin of the exponent $1/4$ observed in Fig.~\ref{fig:square_gapT} at low temperatures, we model the system's dynamics by a simple classical oscillation in a quartic potential and in thermal equilibrium, described by the Hamiltonian 
\begin{equation}
    H_4 = \frac{p^2}{2m} + \lambda x^4,
    \label{eq:H4_no_entropy}
\end{equation}
where the first term is an effective kinetic term. 
Then, by a straightforward dimensional analysis, one finds that the frequency of the oscillation scales as $\omega \sim \lambda^{1/4} E^{1/4}$ for an excitation of energy $E$. 
Associating the frequency with the gap, $\Delta \sim \omega$, and taking into account $E\sim T$, we obtain $\Delta \sim T^{1/4}$, explaining our numerical observations (see End Matter for an alternative numerical approach to the $T^{1/4}$ gap scaling). 
The effective Hamiltonian $H_4$ accounts only for the \emph{energetic} contribution of the quartic mode, and is therefore expected to be accurate only at very low temperatures. 

\textit{Entropic effects.}--- At moderate finite temperatures, the entropic effects must be considered.
These contributions can be effectively modeled by including an additional $\sim Tx^2$ term in  Eq.~\eqref{eq:H4_no_entropy},  which arises from integrating out other spin-wave modes that interact with the quartic mode. This leads to a minimal modified effective Hamiltonian
\begin{equation}
    H_{\rm eff}= H_4 + \alpha T x^2 = \frac{p^2}{2m} + \lambda x^4 + \alpha T x^2.
    \label{eq:quartic-osci-ent}
\end{equation}
The additional term is negligible at very low temperatures, but it becomes comparable to, and ultimately dominates over the quartic term as temperature increases.
To determine the temperature dependence of the frequency of the oscillator in Eq.~\eqref{eq:quartic-osci-ent}, we perform a mean-field decoupling of the $x^4$ term~\cite{sup}, yielding 
\begin{equation}
    \Delta =\sqrt{\frac{\alpha T}{m}} \left (1+\sqrt{1 + \frac{2\,T^*}{T}} \right )^{1/2},
    \label{eq:omegafun}
\end{equation}
where $T^* \equiv 6\lambda / \alpha^2$. This result has two interesting limits: i) for $T \gg T^*$, where the entropic contribution dominates over the quartic term, $\Delta\sim \sqrt{T}$, and ii) for $T \ll T^*$, where the quartic term remains dominant, $\Delta\sim T^{1/4}$. Thus, the temperature $T^*$ sets a rough crossover energy scale below which the quartic mode gap scales as $T^{1/4}$ and above which it follows $\sqrt{T}$. This result not only explains our observation that the gap increases more rapidly at higher temperatures, but also provides a phenomenological equation that can be fitted to our numerical simulations.

In Fig.~\ref{fig:square_gapT}(b) and \ref{fig:square_gapT}(c), we fit our numerical results for the 2D and 3D models to the prediction of the phenomenological theory of Eq.~(\ref{eq:omegafun})~\cite{sup}. 
For both models, we find a good agreement between the simulations and the phenomenological theory, especially at low temperatures. The region where the phenomenological model predicts a $\sim \sqrt{T}$ behavior is harder to identify, but is still identifiable in our numerical data.
Specifically, the red line in Fig.~\ref{fig:square_gapT}(b) corresponds to a plain $\sim \sqrt{T}$ function demonstrating that in a small intermediate temperature regime, our results indeed follow this behavior approximately. 
The vertical lines in Fig.~\ref{fig:square_gapT}(b) and \ref{fig:square_gapT}(c) show the characteristic crossover temperature $T^*$ obtained from our fits. 

Finally, for both the 2D and 3D models, significant deviations between the simulations and the phenomenological model occur near the critical temperature $T_c$. 
This is expected, since the order parameter fluctuations become strong and the fundamental underlying assumption of a state selection with wave vector $\bm{Q}_{\rm B}$ ceases to be fulfilled. 

\textit{Finite-size effects.}--- As stated previously, finite-size effects become important at low temperatures, which is also evident from the $L$-dependent crossover temperatures $T^*$ highlighted in Fig.~\ref{fig:square_gapT}(b) and \ref{fig:square_gapT}(c). 
To explain this, we formally derive here the phenomenological model of Eq.~\eqref{eq:quartic-osci-ent} presented above based on heuristic arguments. 
The first step consists of writing the spins in terms of canonically conjugate variables $\{x_i,p_j\}=\delta_{ij}$,
\begin{eqnarray}
\bm{S}_{i} &=& \sqrt{1-\frac{1}{4}(x_i^2+p_i^2)}\,x_i \,\vhat{x}_i +
\sqrt{1-\frac{1}{4}(x_i^2+p_i^2)}\,p_i \,\vhat{y}_i \nonumber \\
% &\hspace{ 1.2 in} 
&+& \left[1-\frac{1}{2}(x_i^2+p_i^2)\right]\,\vhat{z}_i,
\label{eq:canon}
\end{eqnarray}
where $\{\vhat{x}_i,\vhat{y}_i,\vhat{z}_i\}$ is any right-handed \emph{local} orthogonal frame. 
This is a faithful representation of spins since $\{x_i,p_j\}=\delta_{ij}$ ensures the canonical Poisson bracket relation for the components of $\bm{S}_i$. 
Inserting Eq.~\eqref{eq:canon} into the spin Hamiltonians above, with the local frame aligned with the ground-state spiral ordering at wave vector $\bm{Q}_{\rm B}$, and Taylor expanding in $x_i$ and $p_i$, allows us to isolate the quartic mode and derive the phenomenological model (see details in End Matter). 
The key finding is that $\lambda$ in Eq.~\eqref{eq:quartic-osci-ent} depends on the system size as $\lambda \sim 1/N$.
This result can be traced back to a prefactor $1/\sqrt{N}$ in the Fourier transform, and implies that the crossover temperature behaves as $T^* \sim 1/N$, which is confirmed by our numerical results [inset of Fig.~\ref{fig:square_gapT}(b)].
This also affects the low-temperature regime, where the gap behaves as $\Delta \sim (T/N)^{1/4}$. 

This result not only explains our observations in the numerical calculations, but also provides important non-trivial information about the approach to the thermodynamic limit $N\to \infty$.
On one hand, the gap closes in the low-temperature regime, $\Delta \to 0$ for $N\to \infty$. 
At the same time, the crossover temperature $T^*$ vanishes for $N\to \infty$. This implies that only the $\Delta \sim \sqrt{T}$ regime arising from entropic effects survives for large enough systems. 
Interestingly, this is the same result that was previously found for order-by-thermal-disorder cases with exact zero modes arising from continuous accidental degeneracies~\cite{Khatua2023PseudoG}. 

Although the quartic term in the energy is suppressed by a factor of $1/N$ for individual magnon excitations, it can re-emerge in other settings. For example, in the presence of true or accidental degeneracies, one can form wide domain walls connecting states within the degenerate manifold at vanishing (or nearly vanishing) cost~\cite{savary2012}. In our case, however, domain walls between the two symmetry-related incommensurate orders cost $O(J)$ because of the quartic potential. Relatedly, interactions among parametrically pumped magnons in ferromagnets can overcome the same $1/N$ suppression when the mode population is large, leading to a range of nonlinear dynamical phenomena~\cite{rezende2020fundamentals}.

\textit{Conclusion.}--- In this Letter, we demonstrated that quartic potentials which generically arise in isotropic spiral spin systems lead to a distinctive spin-wave gap of the form $\Delta\sim \sqrt{T}$ in the thermodynamic limit. 
Interestingly, this gap scaling does not result directly from spin oscillations in the quartic potential, but rather from entropic effects that consider the interaction between the quartic mode and the other (macroscopic number of) harmonic modes. 
The absence of direct signatures of quartic oscillation (which lead to a gap scaling $\Delta\sim T^{1/4}$) in the thermodynamic limit is a subtle finite-size effect, as we showed using analytical and numerical methods.

The firm understanding of the role of quartic oscillations developed here for classical spin systems paves the way for future investigations of similar quantum effects at zero and nonzero temperature~\cite{Hickey25}. 
By being controlled by the inverse spin length $1/S$, such effects are expected to conspire with thermal fluctuations to again generate dynamical spin wave gaps $\Delta$. 
The ultimate goal of such studies will be to investigate the impact of finite spin lengths $1/S>0$ on the temperature dependence $\Delta(T)$ in isotropic spiral spin systems and, in a parallel pursuit, seek to expose such behavior in real candidate materials such as ZnCr$_2$Se$_4$~\cite{Inosov2020}.

\textit{Acknowledgments.}--- We thank D. Peets and D. Inosov for helpful discussions. A. F. and M. G. G. acknowledge the use of the JUWELS cluster at the Forschungszentrum J{\"u}lich and the HPC Service of ZEDAT, Freie Universit{\"a}t Berlin. S. K. acknowledges financial support from the Deutsche Forschungsgemeinschaft (DFG, German Research Foundation) under Germany’s Excellence Strategy through the W\"urzburg-Dresden Cluster of Excellence on Complexity and Topology in Quantum Matter -- ct.qmat (EXC 2147, project-id 390858490). The work at the U. of Waterloo and the U. of Windsor was supported by the NSERC of Canada (M. J. P. G. and J. G. R.). M. J. P. G. acknowledges support from the Canada Research Chair program (Tier I).

%\bibliography{biblio.bib}
\input{main.bbl}

%----------------------------------------------------------
%------------------ End Matter ----------------------------
%----------------------------------------------------------

\onecolumngrid

\vspace{\columnsep}

\begin{center}

\textbf{End Matter}

\end{center}

\vspace{\columnsep}

\twocolumngrid
\renewcommand{\theequation}{E\arabic{equation}}
\renewcommand{\thefigure}{E\arabic{figure}}
\setcounter{equation}{0}
\setcounter{figure}{0}
\appendix

\begin{figure}[t!]
    \centering
    \includegraphics[width=0.9\linewidth]{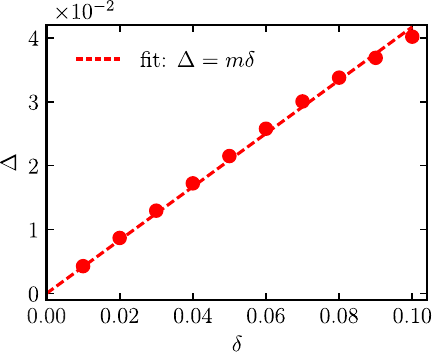}
    \caption{Gap computed as the maximum of $\mathcal{S}(\bm{Q}_{\rm aB},\omega)$ as a function of the perturbation strength $\delta$ for the 2D model. The data are fitted with a linear function ($m \approx 0.4$).}
    \label{fig:square_delta_pert}
\end{figure}

\textit{Section A: Numerical simulation of the $T^{1/4}$ behavior.}---
The quartic perturbation in Eq.~\eqref{eq:quartic} is energetically softer than all other possible deformations of the ground state (apart from global spin rotations).
Therefore, we can alternatively access the low-temperature behavior by initializing the system precisely in the exact perturbed state described by Eq.~\eqref{eq:quartic}, and then obtain its dynamics via sMD for varying perturbation strengths $\delta$. 
The sMD results for the 2D model then show that the dynamical spin structure factor $\mathcal{S}(\bm{Q}_{\rm aB}, \omega)$ at the anti-Bragg point is gapped and the gap $\Delta$ scales linearly with the perturbation strength, $\Delta \sim \delta$ (see Fig.~\ref{fig:square_delta_pert}). 
Based on this result, we can infer the temperature dependence of the gap by considering that the energy for a quartic mode is $E \sim \delta^4$, and $E \sim T$ by the principle of equipartition. 
Thus, we find that $\delta \sim T^{1/4}$ yields a gap that scales as $\Delta \sim T^{1/4}$ in agreement with the dimensional analysis presented in the main text.

\textit{Section B: Derivation of the phenomenological model.}---
To derive the phenomenological quartic oscillator model from the microscopic spin Hamiltonian, we start from a different representation of the classical spins:
\begin{eqnarray}
\bm{S}_{i} &=& \sqrt{1-\frac{1}{4}(x_i^2+p_i^2)}\,x_i \,\vhat{x}_i +
\sqrt{1-\frac{1}{4}(x_i^2+p_i^2)}\,p_i \,\vhat{y}_i \nonumber\\
&+&\left[1-\frac{1}{2}(x_i^2+p_i^2)\right]\,\vhat{z}_i,
\label{eq:diff-rep}
\end{eqnarray}
where the real variables $x_i$, $p_i$ satisfy the Poisson bracket relation $\{x_i, p_j\} = \delta_{ij}$, and $\{\vhat{x}_i, \vhat{y}_i, \vhat{z}_i\}$ is any right-handed \emph{local} orthogonal coordinate frame~\footnote{This representation is a classical form of the Holstein-Primakoff representation~\cite{Holstein1940} with unit spin length and $a_i = (x_i+ip_i)/\sqrt{2}$.}. 
This representation, together with the Poisson bracket relation of $x_i$ and $p_i$ reproduces the canonical Poisson bracket for spin components $\{S_i^\eta, S_j^\rho\}~=~\delta_{ij}\sum_\sigma \epsilon_{\eta\rho\sigma}S_i^\sigma$. 
We further choose the local frame $\{\vhat{x}_i, \vhat{y}_i, \vhat{z}_i\}$ to be aligned with the ground-state spiral ordering at wave vector $\bm{Q}_{\rm B}$ 
\begin{eqnarray}
    \vhat{x}_i &=& -\sin(\bm{Q}_{\rm B}\cdot\bm{r}_i)\,\vhat{x} + \cos(\bm{Q}_{\rm B}\cdot\bm{r}_i)\,\vhat{y},\nonumber\\
    \vhat{y}_i &=& \vhat{z},\nonumber\\
\vhat{z}_i &=& \cos(\bm{Q}_{\rm B}\cdot\bm{r}_i)\,\vhat{x} + \sin(\bm{Q}_{\rm B}\cdot\bm{r}_i)\,\vhat{y},
\label{eq:loc-frame}
\end{eqnarray}
where $\{\vhat{x}, \vhat{y}, \vhat{z}\}$ are the Cartesian directions in  the \emph{global} coordinate frame.  
Substituting Eqs.~\eqref{eq:diff-rep} and \eqref{eq:loc-frame} into the microscopic spin Hamiltonians, Taylor expanding in $x_i$, $p_i$, and keeping terms up to quadratic order, yields the energy
\begin{equation}
    H_2 = \frac{1}{2}\sum_{i,\nu} J_{\nu}\left[\cos(\bm{Q}_{\rm B} \cdot \bm{\nu})\left(x_ix_{i+\nu} - x_i^2 - p_i^2\right) + p_i p_{i+\nu}\right],
\end{equation}
where $J_\nu$ denotes the coupling strength on the bond $\bm{\nu}$ from the lattice site $i$. 
Further, considering the Fourier transform 
\begin{equation}
    x_i = \frac{1}{\sqrt{N}}\sum_{\bm{k}}e^{i\bm{k}\cdot\bm{r}_i}x_{\bm{k}}, \hspace{0.5 cm} p_i = \frac{1}{\sqrt{N}}\sum_{\bm{k}}e^{i\bm{k}\cdot\bm{r}_i}p_{\bm{k}}, 
\end{equation}
with $N$ being the number of spins in the system, we obtain
\begin{equation}
    H_2 = \frac{1}{2}\sum_{\bm{k}} \left[L_{\bm{k}}\, x_{-\bm{k}}x_{\bm{k}} + M_{\bm{k}}\, p_{-\bm{k}}p_{\bm{k}}\right],
    \label{eq:SHO}
\end{equation}
where 
\begin{eqnarray}
   L_{\bm{k}} &=& \sum_{\nu} J_{\nu}\cos(\bm{Q}_{\rm B}\cdot\bm{\nu})\left(e^{i\bm{k}\cdot\bm{\nu}} - 1\right),\nonumber\\ 
    M_{\bm{k}} &=& \sum_{\nu} J_{\nu}\left(e^{i\bm{k}\cdot\bm{\nu}} - \cos(\bm{Q}_{\rm B}\cdot\bm{\nu})\right).
\end{eqnarray}
In reciprocal space, we have the Poisson bracket $\{x_{\bm{k}},p_{\bm{k}'}\} = \delta_{\bm{k}+\bm{k}'}$ which thus implies the conjugate variable to $x_{\bm{k}}$ is $p_{-\bm{k}}$. 
Therefore, the quadratic Hamiltonian in Eq.~\eqref{eq:SHO} describes a simple harmonic oscillator with frequency $\omega_{\bm{k}} = \sqrt{L_{\bm{k}}M_{\bm{k}}}$. 
Thus, a zero mode occurs when $L_{\bm{k}}$ = 0, $M_{\bm{k}} = 0$, or both are zero. 
We find $L_{\bm{0}} = 0$ and $M_{\bm{Q}_{\rm B}}= M_{\bm{Q}_{\rm aB}}= 0$. 
To focus on the mode at $\bm{Q}_{\rm aB}$, we retain only the wave vectors $\bm{Q}_{\rm aB}$ and $-\bm{Q}_{\rm aB}$ in the Fourier expansion (as conjugate variables appear at $\bm{k}$ and $-\bm{k}$): 
\begin{eqnarray}
    x_i = \frac{1}{\sqrt{N}}\left(e^{i\bm{Q}_{\rm aB}\cdot\bm{r}_i}x_{\bm{Q}_{\rm aB}} + e^{-i\bm{Q}_{\rm aB}\cdot\bm{r}_i}x_{-\bm{Q}_{\rm aB}} \right),\nonumber\\ 
    p_i = \frac{1}{\sqrt{N}}\left(e^{i\bm{Q}_{\rm aB}\cdot\bm{r}_i}p_{\bm{Q}_{\rm aB}} + e^{-i\bm{Q}_{\rm aB}\cdot\bm{r}_i}p_{-\bm{Q}_{\rm aB}} \right).
    \label{eq:aB-mode}
\end{eqnarray}
The real valuedness of $x_i$ and $p_i$ implies the complex conjugate $x^*_{-\bm{k}} = x_{\bm{k}}$ and $p^*_{-\bm{k}} = p_{\bm{k}}$ for any $\bm{k}$. 
Making use of these conditions and the relation  $\{x_{\bm{k}},p_{\bm{k}'}\} = \delta_{\bm{k}+\bm{k}'}$, we rewrite the Fourier components as the following 
\begin{eqnarray}
    x_{\bm{Q}_{\rm aB}} &=& (X + i\tilde{P})/\sqrt{2}, \hspace{0.2 cm} x_{-\bm{Q}_{\rm aB}} = (X - i\tilde{P})/\sqrt{2}, \nonumber\\
     p_{\bm{Q}_{\rm aB}} &=& (P - i\tilde{X})/\sqrt{2}, \hspace{0.2 cm} p_{-\bm{Q}_{\rm aB}} = (P + i\tilde{X})/\sqrt{2},
\end{eqnarray}
where $X,P$ and $\tilde{X}, \tilde{P}$ satisfy the Poisson bracket relations $\{X,P\} = \{\tilde{X},\tilde{P}\}=1$, and other Poisson brackets are zero. 
Eq.~\eqref{eq:aB-mode} then becomes 
\begin{eqnarray}
    x_i = \sqrt{\frac{2}{N}}\left(X\cos(\bm{Q}_{\rm aB}\cdot\bm{r}_i) -\tilde{P} \sin(\bm{Q}_{\rm aB}\cdot\bm{r}_i) \right),\nonumber\\ 
    p_i = \sqrt{\frac{2}{N}}\left(P\cos(\bm{Q}_{\rm aB}\cdot\bm{r}_i) + \tilde{X}\sin(\bm{Q}_{\rm aB}\cdot\bm{r}_i) \right).
    \label{eq:aB-mode-mod}
\end{eqnarray}
Substituting Eq.~\eqref{eq:aB-mode-mod} into Eq.~\eqref{eq:diff-rep}, Taylor expanding it up to quartic order in $X,P,\tilde{X},\tilde{P}$, and then inserting the expansion in the microscopic spin Hamiltonian yields an effective quartic oscillator Hamiltonian, which includes several quartic combinations of those four variables. 
Most importantly, these quartic terms in the Hamiltonian have an explicit system-size dependence, which is $O(1/N)$; Eq.~\eqref{eq:aB-mode-mod} shows that each quartic term is $O(1/N^2)$, with the microscopic spin Hamiltonian having a sum over bonds, which gives a contribution of $O(N)$ -- combining these contributions, we ultimately get a $1/N$ coefficient for the quartic terms in the final effective quartic Hamiltonian. 
We thus see that the prefactor of the quartic potential term in Eq.~\eqref{eq:quartic-osci-ent} scales as $\lambda \sim 1/N$. 

With the proper $N$-dependent scaling of the effective quartic Hamiltonian in reciprocal space at hand, proceeding next with a mean-field decoupling of the quartic terms as done for Eq.~\eqref{eq:quartic-osci-ent}, yields an effective simple harmonic oscillator whose frequency -- and hence the gap $\Delta$ -- has the same form as in Eq.~\eqref{eq:omegafun}, with the crossover temperature $T^*\sim 1/N$. 
Thus, the gap at low temperatures behaves as $\Delta \sim (T/N)^{1/4}$, while at high temperatures, where the entropic contribution dominates, $\Delta \sim\sqrt{T}$ --- independent of the system size. 
Consequently, in the thermodynamic limit ($N\rightarrow \infty$), $T^*\rightarrow 0$ and only one scaling relation --- $\Delta \sim \sqrt{T}$ --- remains.
\clearpage

%----------------------------------------------------------
%------------------ Supplemental Material -----------------
%----------------------------------------------------------

\beginsupplement

\onecolumngrid

\begin{center}
{\bf \large Supplemental Material for ``Anharmonic Collective Oscillations in Isotropic Spin Systems and their Spectroscopic Signatures''}\\[1.5em]

Anna Fancelli,$^{1,2}$ Mat\'ias G. Gonzalez,$^{1,2,3}$ Subhankar Khatua,$^{4,5,6}$ Bella Lake,$^{1,7}$ Michel J. P. Gingras,$^{4,8}$ Jeffrey G. Rau,$^{5}$ and Johannes Reuther$^{1,2}$\\[0.5em]

\textit{\small
$^1$Helmholtz-Zentrum Berlin f\"ur Materialien und Energie, Hahn-Meitner-Platz 1, 14109 Berlin, Germany\\
$^2$Dahlem Center for Complex Quantum Systems and Fachbereich Physik, Freie Universit\"at Berlin, Arnimallee 14, 14195 Berlin, Germany\looseness=-1\\
$^3$Institute of Physics, University of Bonn, Nussallee 12, 53115 Bonn, Germany\\
$^4$Department of Physics and Astronomy, University of Waterloo, Waterloo, Ontario N2L 3G1, Canada\\
$^5$Department of Physics, University of Windsor, 401 Sunset Avenue, Windsor, Ontario N9B 3P4, Canada\\
$^6$Institute for Theoretical Solid State Physics, IFW Dresden and W\"urzburg-Dresden\\
Cluster of Excellence ct.qmat, Helmholtzstr. 20, 01069 Dresden, Germany\\
$^7$Institut f\"ur Festk\"orperforschung, Technische Universit\"at Berlin, 10623 Berlin, Germany\\
$^8$Waterloo Institute for Nanotechnology, University of Waterloo, Waterloo, Ontario N2L 3G1, Canada
}
\end{center}

\section{Energy expansion of the quartic perturbation}

In this section, we present the series expansion of the energy for a spiral state perturbed by a perpendicular perturbation with wave vector $\bm{Q}_{\rm aB}$ in both the square and cubic lattices, demonstrating the quartic dependence of the energy of such a state on the perturbation amplitude.

\subsection{Square lattice}

To illustrate how quartic modes emerge in a spiral ground state perturbed by a perpendicular perturbation with a wave vector corresponding to the non-selected spiral, $\bm{Q}_{\rm aB}$, we consider a generic spiral with period $L$. 
First, it is useful to rewrite the perturbed state given in Eq. (3) of the main text as:
\begin{equation}
    \bm{S}^{\text{quartic}}_{i} = \frac{1}{\sqrt{1 + \delta^2 \cos^2{\left( \frac{2 \pi}{L}x \right) }}}  \begin{pmatrix}  \cos{\left( \frac{2 \pi}{L}y \right) } \\ \sin{\left( \frac{2 \pi}{L}y \right) } \\  \delta \cos{\left( \frac{2 \pi}{L}x \right) } \end{pmatrix},
    \label{sm_eqsm:quartic_per}
\end{equation}
where the position of site $i$ is given by $\bm{r}_i = (x, y)$ and we set $\bm{Q}_{\rm B}=(0, 2 \pi /L)$, $\bm{Q}_{\rm aB}=( 2 \pi /L,0)$. Furthermore, we have set the arbitrary phases to be $\phi = \phi^\prime=0$.
Next, consider the energy term associated with the $n$th nearest-neighbor interactions:
\begin{equation}
    J_n \sum_{i} \sum_{j\in n^{\text{th}}{\text{ n.n of }}i }  \bm{S}_i \cdot \bm{S}_j .
    \label{sm_eqsm:Jn}
\end{equation}
This term is proportional to the scalar product between the spin at site $\bm{r}_i = (x, y)$ and its $n^{\textrm{th}}$ nearest neighbors, given by $\bm{r}_j = (x \pm a, y \pm b)$ and $(x \pm b, y \pm a)$. 
To avoid double counting, we include only one term from each pair related by a global minus sign: $(x + a, y + b)$, $(x - a, y + b)$, $(x - b, y + a)$, and $(x - b, y - a)$. 
These four terms form two pairs related by a $90^\circ$ rotational symmetry around the $\bm{\hat{z}}$ axis. 
To show that the quadratic term vanishes, it is sufficient to consider one such pair, summed over the entire lattice:
\begin{equation}
   \sum_{x,y}  \bm{S}(x, y) \cdot \bm{S}(x + a, y + b)+ \bm{S}(x, y) \cdot \bm{S}(x - b, y + a).
\end{equation}
Expanding the scalar products in a series of $\delta$, we obtain for the second-order terms the following expression:
\begin{equation}
\begin{aligned}
     \sum_{x,y}  &  \delta^2 \cos{\left( \frac{2 \pi}{L}x \right) } \left[ \cos{\left( \frac{2 \pi}{L} (x+a) \right)} +\cos{\left( \frac{2 \pi}{L} (x-b) \right)}  \right]  \\
      & - \frac{\delta^2}{2} \cos{\left( \frac{2 \pi}{L}b \right) } \left[ \cos^2{\left( \frac{2 \pi}{L} (x+a) \right)} +\cos^2{\left( \frac{2 \pi}{L}x \right)}   \right]  \\
      &  - \frac{\delta^2}{2} \cos{\left( \frac{2 \pi}{L}a \right) } \left[ \cos^2{\left( \frac{2 \pi}{L} (x-b) \right)} +\cos^2{\left( \frac{2 \pi}{L}x \right)}   \right]. 
\end{aligned}
\end{equation}
Using algebraic manipulations and the fact that summing over the entire lattice causes terms proportional to $ \cos{\left( \frac{2 \pi}{L}x \right)}$ and $ \sin{\left( \frac{2 \pi}{L}x \right)}$, or equivalent expressions for $y$, to vanish, we arrive at:
\begin{equation}
    \sum_{x,y} \frac{\delta^2}{2} \cos{\left( \frac{4 \pi}{L}x \right) } \left[ \cos{\left( \frac{2 \pi}{L}b \right)} \sin^2{\left( \frac{2 \pi}{L}a \right)} + \sin{\left( \frac{2 \pi}{L}b \right)} \cos^2{\left( \frac{2 \pi}{L}a \right)}  \right] = 0.
\end{equation}
This final expression vanishes when summing over the entire lattice. 
Therefore, while the existence of a spiral ground state depends on the Hamiltonian, the quartic nature of the mode involving a perpendicular perturbation with the wave vector of the non-selected spiral is a general property of the spiral ground state, independent of the specific Hamiltonian. The same argument also applies to two-dimensional spirals.

The energy expansion up to the fourth order for the state in Eq. \eqref{sm_eqsm:quartic_per}, specific to the square lattice discussed in the main text, is given by
\begin{equation}
    e(\delta)=\sum_{i=1}^3 e_i(\delta) \approx -1.00729 + 0.03906 \delta^4+ O(\delta^6),
\end{equation} 
where $e_i$ with $i=1$, 2, 3 are the energy contributions from the different lattice bonds with couplings $J_1$, $J_2$, $J_3$, respectively.

\subsection{Cubic lattice}

To compute the energy expansion, we consider a state similar to the one in Eq. \eqref{sm_eqsm:quartic_per}, generalized to the equal superposition of the two quartic perturbations $S_i^z \sim \delta[\cos(\bm{Q}_{\rm aB_1}\cdot \bm{r}_{i}) + \cos(\bm{Q}_{\rm aB_2}\cdot \bm{r}_{i})]$. 
We rewrite the state as:
\begin{equation}
    \bm{S}^{\text{quartic}}_{i}=\frac{1}{\sqrt{1+ \delta^2( \cos^2{(\frac{2\pi}{L}x)}+\cos^2{(\frac{2\pi}{L}z)}) }} \begin{pmatrix}  \cos{(\frac{2\pi}{L}y)} \\\sin{(\frac{2\pi}{L}y)} \\  \delta( \cos{(\frac{2\pi}{L}x)}+\cos{(\frac{2\pi}{L}z)})\end{pmatrix} ,
    \label{sm_eqsm:quartic_pert_cubic}
\end{equation}
where $\bm{r}_i=(x,y,z)$, $\bm{Q}_{\rm B}=(0,\frac{2\pi}{L},0)$, $\bm{Q}_{\rm aB_1}=(\frac{2\pi}{L},0,0)$, and $\bm{Q}_{\rm aB_2}=(0,0,\frac{2\pi}{L})$. 
To show that the terms proportional to $\delta^2$ in the expression in Eq. \eqref{sm_eqsm:Jn} vanish, the procedure is analogous to the one presented for the square lattice.
The only difference in the cubic lattice case is that we must consider three contributions, as the perturbation consists of two terms. 
Specifically, the scalar product is taken between the spin at site $\bm{r}_i=(x,y,z)$ and three of its $n^{\rm th}$ nearest neighbors: (1) the spin at $(x+a,y+b,z+c)$ (2) the spin at $(x-b,y+a,z+c)$ obtained by a $90^\circ$ rotation around the $\hat{\bm{z}}$-axis (3) the spin at $(x+a,y-c,z+b)$ obtained by a $90^\circ$ rotation around the $\hat{\bm{x}}$-axis.

The energy expansion up to the fourth order for the state in Eq. \eqref{sm_eqsm:quartic_pert_cubic}, specific to the cubic lattice discussed in the main text, is given by
\begin{equation}
    e(\delta)=e_1(\delta)+e_2(\delta)+e_4(\delta) \approx -1.48114 + 0.153863 \delta^4+ O(\delta^6).
\end{equation} 

\section{Details on the Monte Carlo simulations}

To generate the spin configurations needed for the spin Molecular Dynamics (sMD) simulations, we perform classical Monte Carlo (cMC) calculations on the square and cubic lattices. 
We use a particular version of the single-spin update algorithm called the adaptive Gaussian step algorithm, which allows for a controlled and fixed $50\%$ acceptance ratio at all temperatures~\cite{Alzate19sm}. Additionally, for each spin-update trial, we perform two over-relaxation steps. Then, at a given temperature, we perform $5\times 10^5$ cMC steps consisting each of $N$ single-spin update trials and $2N$ over-relaxation steps, where $N$ is the number of spins in the system. Data for measurements is collected during the second half of the cMC steps.

For both lattices, we perform two types of simulations: cooling down and heating, both with logarithmic temperature steps. 
For the cooling-down simulations, we do 120 temperature steps from $T/J = 2$ down to $T/J=0.01$, where $J$ is the largest coupling. 
Then we perform an extra 31 steps down to $T/J=10^{-12}$. 
For the heating simulations, we depart from a ground-state configuration and perform 34 logarithmic steps from $T/J = 10^{-12}$ up to $10^{-1}$ (16 steps in the cubic case).
We save the last configuration at each temperature as a benchmark point.

From each configuration at a desired temperature, we generate 120 new spin configurations for the Molecular Dynamics simulations. 
This is done by performing $5\times 10^5$ cMC steps between spin configurations, implying 60 million cMC steps between the first and last spin configurations at a given temperature. 

In the case of the square lattice, we do this for $N=L^2$ lattices with $L=20$, 40, 60, and 80. In the case of the cubic lattice, we do this for $N=L^3$ lattices with $L=5$, 10, 15, and 20.
In both cases, we use periodical boundary conditions, which do not frustrate the 5-site spiral ground state. 

\begin{figure}[h!]
    \centering
    \includegraphics[width=0.9\linewidth]{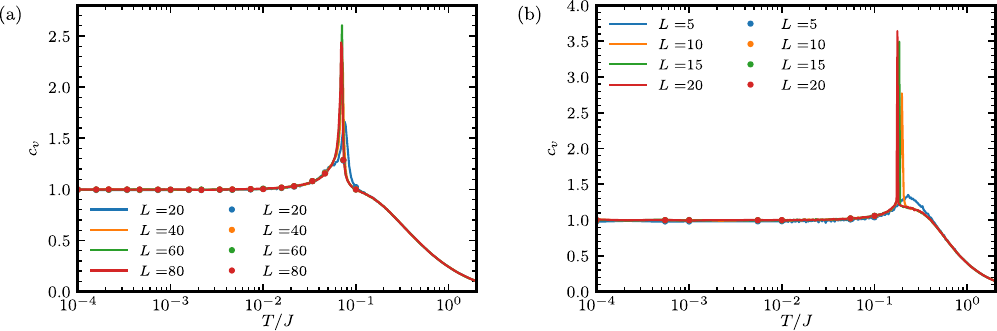}
    \caption{Classical Monte Carlo calculations for the specific heat $c_v(T)$. The data from the cooling down simulations is shown in a continuous line, while the results from heating up are shown by the full dots. The temperature is shown in $T/J$, where $J$ is the largest coupling in each case. Panels (a) and (b) correspond to the square and cubic lattice models, respectively. }
    \label{sm_fig:figcmc1}
\end{figure}

For completeness, we show in Fig.~\ref{sm_fig:figcmc1} the specific heat $c_v(T)$ from the cMC calculations; in panel (a) for the square lattice model and in panel (b) for the cubic lattice. 
In both cases, the lines indicate the calculations for the cooling down simulations, while the dots are the results for the heating up simulations, noting an excellent agreement between both. 
Also, both calculations find evidence for only a single phase transition into a spiral phase. 

\section{Details on the spin Molecular Dynamics simulations}

In this section, we explain how the spin Molecular Dynamics (sMD) simulations are performed and how the dynamical structure factor is computed within this framework.

\subsection{Overview of the method and numerical details}

The sMD simulations consist of numerically integrating the classical equations of motion for the spins:
\begin{equation}
    \frac{d\bm{S}_i}{dt} = \bm{S}_i \times \bm{h}_i,
    \label{sm_eqsm:motion_h}
\end{equation}
where $\bm{h}_i=-\frac{\partial H}{\partial \bm{S}_i }$ is the local effective field acting on $\bm{S}_i$, arising from interactions with neighboring spins. These equations correspond to the Landau-Lifshitz equations without the damping term \cite{landau1935theorysm}. The numerical integration is carried out using the Runge-Kutta Cash-Karp method, implemented through the Boost C++ Library \cite{ahnert2012boostsm}.

\subsection{Computing the dynamical structure factor}

The central quantity of our study is the dynamical structure factor 
\begin{equation}
    \mathcal{S}(\bm{q},\omega) = \frac{1}{2\pi N}  \sum_{i,j=1}^N\int_{-\infty}^{\infty} e^{i\omega t} e^{-i \bm{q} (\bm{r}_i-\bm{r}_j)} \langle \bm{S}{_i}(0) \cdot  \bm{S}{_j}(t) \rangle dt
    \label{sm_eqsm:dsf}
\end{equation}
evaluated at $\bm{q}=\bm{Q}_{\rm aB}$.
To compute $\mathcal{S}(\bm{Q}_{\rm aB},\omega)$, we follow the approach of Ref. \cite{Zhang2019Dynamicalstructsm}. 
Specifically, the spin Fourier transform for each spin component is computed during the simulation. 
At the end of the time evolution, $\mathcal{S}(\bm{Q}_{\rm aB},\omega)$ is obtained by multiplying the Fourier transform by its complex conjugate and summing over the components. 
For the calculation of $ \mathcal{S}(\bm{Q}_{\rm aB},\omega)$ at finite temperature, the results are obtained by averaging over 120 simulations.
Each simulation starts from an independent, equilibrated MC configuration, which is obtained by heating up the system starting from zero temperature.

\section{Fitting procedures}

In this section, we describe the fitting procedures used in this work. First, we explain how the gap value is extracted from the dynamical structure factor. Then, we discuss how these extracted values are fitted using the function in Eq. (6) of the main text, which is derived from the phenomenological model. The results of the second fit are shown in Figs. 2(b) and 2(c).

\subsection{Gap extraction from the dynamical structure factor}

The value of the gap at $\bm{q}=\bm{Q}_{\rm aB}$ is obtained by fitting $\mathcal{S}(\bm{Q}_{\rm aB},\omega)$ with a Lorentzian function multiplied by a scaling factor $A$:
\begin{equation}
    f(x) = A \frac{\gamma}{\pi [\gamma^2 + (x-x_0)^2]}.
    \label{sm_eqsm:lor}
\end{equation}
The value of the gap corresponds to the position of the maximum, given by $x_0$. 
For each linear size and temperature, we compute the difference between the gap value extracted from the fit and the position of the maximum of $\mathcal{S}(\bm{Q}_{\rm aB},\omega)$. Since all observed differences lie within $3\gamma/4$ (square lattice) and $\gamma/2$ (cubic lattice), we adopt these values as estimates of the error.

\subsection{Fitting the simulated gaps}

\begin{figure}[h!]
    \centering
    \includegraphics[width=0.9\linewidth]{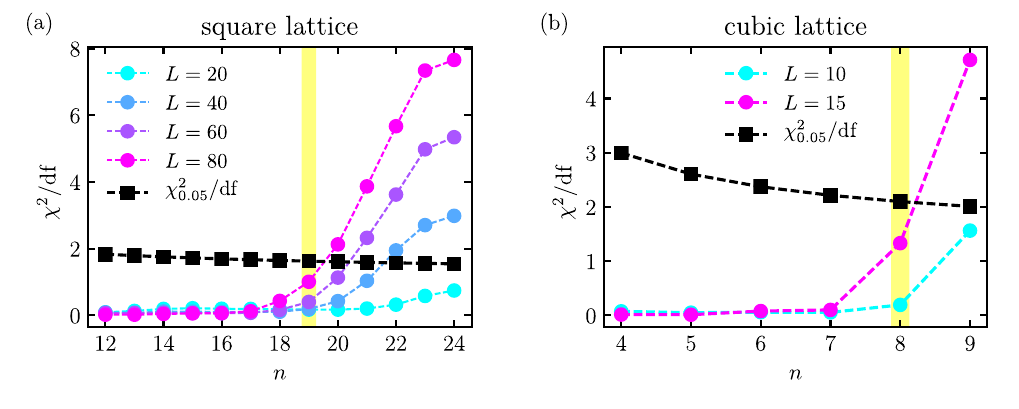}
    \caption{(a) The reduced chi-squared per degree of freedom, $\chi^2/\text{df}$, as a function of $n$, the number of points used to fit the gap values with the function in Eq. (6), for the square lattice with $L = 20, 40, 60, 80$.
    Since this function depends on two parameters, the degrees of freedom are given by $n - 2$.
    The black squares represent the critical value $\chi_{0.05}^2/\text{df}$.
    The yellow region highlights the maximum number of points for which $\chi^2/\text{df}$ remains below the critical chi-squared value for all system sizes.
    These selected points, starting from the lowest-temperature point, are those used in the fit shown in Fig. 2(b) of the main text. 
    (b) Same as (a) for the cubic lattice, with $L = 10,15$.}
    \label{sm_fig:chisquared}
\end{figure}

The gap values are fitted using the function in Eq.~(6) of the main text, which describes the gap behavior derived from the phenomenological model.
At low temperatures, this function predicts $\Delta \sim T^{1/4}$. 
This scaling is confirmed by the linear fit of $\log{T}$ versus $\log{\Delta}$, shown by the blue line in Fig.~2(b) for the square lattice and the blue line in Fig.~2(c) for the cubic lattice.
Thus, we verify that the function in Eq.~(6) provides a good fit to the data at low temperatures.
Starting from the lowest temperatures, we gradually increase the number of data points used in the fit. 
For each step, we compute $\chi^2/\text{df}$, where $\text{df} = n - 2$ is the number of degrees of freedom (with $n$ being the number of points used, minus two since the fit depends on two parameters).
We then compare $\chi^2/\text{df}$ with the chi-squared critical value at a 95\% confidence level, denoted as $\chi_{0.05}^2/\text{df}$. 
The computed $\chi^2/\text{df}$ values and corresponding critical values for the two lattices are shown in Fig. \ref{sm_fig:chisquared}.
The fits presented in Figs.~2(b) and 2(c) correspond to the maximum number of points where $\chi^2/\text{df}$ remains below the critical threshold across all linear system sizes. 
This results in $n=19$ for the square lattice and $n=8$ for the cubic lattice.

\section{Additional data for the cubic lattice}

In this section, we present additional data for the cubic lattice. 
Specifically, the energy gap as a function of the perturbation strength, and the gap as a function of temperature at the two anti-Bragg wave vectors.

\subsection{Dynamical gap as a function of the perturbation strength }

\begin{figure}[h!]
    \centering
    \includegraphics[width=0.9\linewidth]{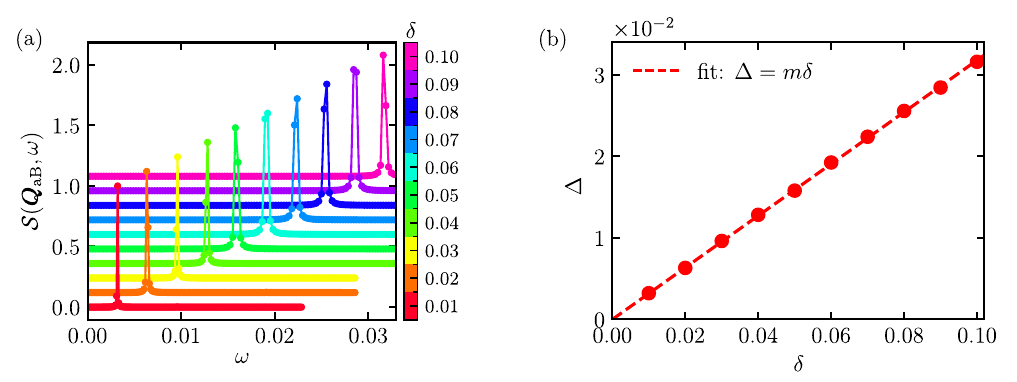}
    \caption{(a) Normalized dynamical structure factor $\mathcal{S}(\bm{q},\omega)$ at $\bm{q} = \bm{Q}_{\rm aB}$ for different values of the perturbation strength $\delta$, obtained from sMD simulations for $L = 15$. 
    The calculation is performed at both $\bm{Q}_{\rm aB_1}$ and $\bm{Q}_{\rm aB_2}$, and since the resulting curves are identical, we use the notation $\bm{Q}_{\rm aB}$ in the plot to represent both points. 
    (b) Gap $\Delta$ at $\bm{q} = \bm{Q}_{\rm aB_1}$, equal to the one at $\bm{q} = \bm{Q}_{\rm aB_2}$,  as a function of the perturbation strength $\delta$. 
    The data are fitted with a linear function ($m \approx 0.32$). }
    \label{sm_fig:fixpert_cubic}
\end{figure}

Similarly to the square lattice discussed in the End Matter, we compute $\mathcal{S}(\bm{Q}_{\rm aB},\omega)$, starting from a configuration in which the quartic modes are excited directly, using the expression in Eq.~\eqref{sm_eqsm:quartic_pert_cubic}. 
The sMD simulations show that $\mathcal{S}(\bm{Q}_{\rm aB_1},\omega)$ and $\mathcal{S}(\bm{Q}_{\rm aB_2},\omega)$ display a peak at the same frequency, that scales linearly with $\delta$, see Fig.~\ref{sm_fig:fixpert_cubic}.

\subsection{Equivalence of the gap at the anti-Bragg points}

\begin{figure}[h!]
    \centering
    \includegraphics[width=0.5\linewidth]{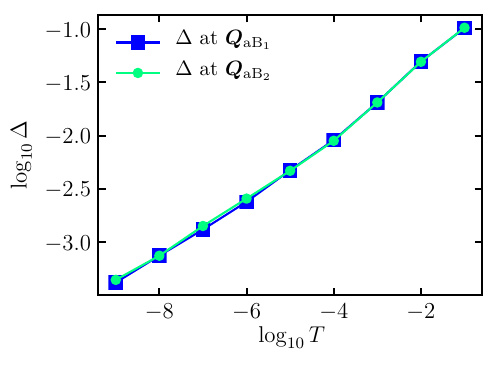}
    \caption{Values of the gap $\Delta$ as a function of temperature, extracted respectively from $\mathcal{S}(\bm{Q}_{\rm aB_1},\omega)$ and $\mathcal{S}(\bm{Q}_{\rm aB_2},\omega)$, for the cubic system with linear size $L=15$. }
    \label{sm_fig:ab1ab2_cubic}
\end{figure}

As explained in the main text, the cubic lattice exhibits quartic modes at the two not-selected spiral wave vectors denoted as $\bm{Q}_{\rm aB_1}$ and $\bm{Q}_{\rm aB_2}$.
As shown in Fig. \ref{sm_fig:ab1ab2_cubic}, the gap at $\bm{Q}_{\rm aB_1}$ and $\bm{Q}_{\rm aB_2}$ is the same, which agrees with the equivalence of these points. 
This agreement further confirms that the initial configurations are equilibrated correctly.
In Fig.~2(c) of the main text, the gap values are obtained from the average of the gaps at the two anti-Bragg points.

\section{Frequency of the quartic oscillator in the presence of the entropic term}
We consider a phenomenological model of a quartic oscillator with an additional entropic contribution, described by the effective Hamiltonian 
\begin{equation}
    H_{\rm eff}= \frac{p^2}{2m} + \lambda x^4 + \alpha T x^2,
    \label{eqsm:quartic-osci-ent}
\end{equation}
where the first two terms are energetic contributions and the last term arises from entropy (see the main text). 
To estimate the frequency of this oscillator, we apply a mean-field decoupling of the quartic term as $x^4 \approx 6\,\avg{x^2}\, x^2$, where $\avg{\cdots}$ denotes thermal average, assuming thermal equilibrium at temperature $T$. 
The factor of $6$ comes from the number of ways to contract $x^4$ into a product involving $\avg{x^2}\,x^2$. 
This decoupling leads to an effective quadratic Hamiltonian
\begin{equation}
    H_{\rm{quad}}\approx \frac{p^2}{2m} + \left(6\lambda \avg{x^2} + \alpha T\right) x^2.
\end{equation}
For such a quadratic Hamiltonian, the thermal average 
\begin{equation}\label{eqsm:freq}
\avg{x^2} = \frac{T}{m\omega^2} = \frac{T}{2\left(6\lambda\avg{x^2} +\alpha T\right)}.
\end{equation}
This equation can be solved self-consistently for $\avg{x^2}$ to find the frequency $\omega$ or, alternatively, one may rewrite Eq.~\eqref{eqsm:freq} as 
\begin{equation}
    \frac{m\omega^2}{2} = \frac{6\lambda T}{m\omega^2} +\alpha T.
\end{equation}
This is a quadratic equation in $\omega$, giving 
\begin{equation}
    \omega =\sqrt{\frac{\alpha T}{m}} \left (1+\sqrt{1 + \frac{2\,T^*}{T}} \right )^{1/2},
\end{equation}
where $T^* = 6\lambda/\alpha^2$. 
This frequency is identified as the thermal gap $\Delta$ of the quartic mode at temperature $T$.

%\bibliography{SM/biblio2.bib}
\input{sm.bbl}

\end{document}